\documentclass{article}
\usepackage[utf8]{inputenc}
\usepackage{graphicx}
\usepackage{pdfpages}
\usepackage{latexsym}
\usepackage{amsbsy}
\usepackage{amsmath}
\usepackage{amssymb}
\usepackage{lineno}
\usepackage{multirow}
\usepackage{url}
\usepackage[hidelinks]{hyperref}
\usepackage{authblk}
\usepackage{doi}
\begin{document}

\date{}

\title{Quantum--classical crossover in finite spin-$\tfrac12$ rings with Dzyaloshinsky--Moriya interaction}

\author[1]{Ra\'ul S\'anchez Gal\'an}
\author[2,3]{Robert Wieser}

\affil[1]{\small 4i Intelligent Insights, 41092
Sevilla, Spain}
\affil[2]{\small School of Physics and Optoelectronic Engineering, Nanjing University of Information Science and Technology, Nanjing 210044, China}
\affil[3]{\small Jiangsu Key Laboratory for Optoelectronic Detection of Atmosphere and Ocean, Nanjing University of Information Science and Technology, Nanjing 210044, China}

\maketitle

\begin{abstract}
We study finite spin-$\tfrac12$ rings with nearest-neighbor Heisenberg
exchange, Dzyaloshinsky--Moriya interaction, and an external magnetic field.
We introduce a phenomenological interpolation parameter between the fully quantum Hamiltonian
and a state-dependent mean-field description. Using dissipative
Gisin--Schr\"odinger dynamics, we analyze the resulting quantum--classical
crossover through local magnetization, connected spin correlations,
single-site entropy, and the saturation field of the fully polarized state.
\end{abstract}


\section{Introduction}

Finite quantum spin systems provide a natural setting in which to study
the crossover between microscopic quantum correlations and effective
classical spin dynamics. Among them, spin rings are particularly useful:
they allow detailed analytical and numerical study while retaining
finite-size effects, frustration, and magnetic-field-induced transitions.
Molecular nanomagnets and engineered spin structures provide physical
realizations of finite spin clusters and have motivated extensive work
on quantum magnetism, molecular spintronics, and nanoscale magnetic
dynamics
\cite{boganiNatMat08,gaitaarino19,morenopinedaNatRevPhys21}.

Earlier work has established the importance of ground-state quantum
numbers, frustration, and coherent dynamics in finite spin rings
\cite{baerwinkelPRB03,schnackJMMM06}.
More recently, odd and even ferric wheels have been compared
experimentally \cite{cutlerAdvSci23}, and spin-$\tfrac12$ molecular rings
with five and six sites have been investigated
\cite{liJACS25,liAngew26}.
These developments motivate the study of how quantum correlations
modify the description of small magnetic rings in terms of local spin
vectors.

The basic microscopic interaction considered here is the Heisenberg
exchange. The spin-$\tfrac12$ Heisenberg chain is a paradigmatic
many-body model whose thermodynamic properties are well understood
through exact and field-theoretic methods
\cite{Auerbach1994,Bethe1931,Hulthen1938,Takahashi1999}.
Finite rings exhibit additional effects arising from discrete momenta
and periodic boundary conditions. In particular, an antiferromagnetic
ring with an odd number of sites cannot support a perfect two-sublattice
N\'eel configuration. This geometrical frustration contributes to
characteristic even--odd differences in its low-energy states and
magnetic response
\cite{baerwinkelPRB03,moessnerPT06,cutlerAdvSci23}.
Even--odd effects have also been studied in short open antiferromagnetic
chains \cite{machensPRB13}.

When spin--orbit coupling is present and bond inversion symmetry is
absent, the symmetric Heisenberg exchange can be supplemented by the
antisymmetric Dzyaloshinsky--Moriya (DM) interaction
\cite{dzyaloshinkyJPCS58,Moriya1960}.
This interaction favors chiral spin arrangements and plays an important
role in spin spirals, skyrmions, and chirality-based information transfer
\cite{Fert2017,heidePRB08,menzelPRL12}.
The DM interaction is often weaker than the Heisenberg exchange
\cite{Moriya1960,Fert2017}; for example, the analysis of Cu benzoate
suggests a staggered DM coupling of order $0.1J$ \cite{Affleck1999}.
Motivated by this scale, we use $D=0.2J$ as an illustrative moderate
coupling in our uniform-DM ring model.
The DM-induced momentum shift and its connection to spiral correlations
were already identified in early studies of one-dimensional spin
systems \cite{KontorovichTsukernik1967}.
For uniform DM exchange, a site-dependent spin rotation removes the
phase from the bulk transverse interaction, producing an XXZ-type
Hamiltonian with a generally twisted boundary condition on a periodic
ring \cite{PerkCapel1976}.
The interplay between this twist and the discrete allowed momenta
affects the finite-size saturation field.

Classical magnetic dynamics with damping is commonly described by the
Landau--Lifshitz--Gilbert equation, in which magnetic moments evolve as
vectors of fixed length in an effective field
\cite{landauPZS35,gilbertIEEE04}.
Quantum spin dynamics additionally permits entanglement and changes in
the magnitudes of local spin expectation values.
Nonlinear dissipative extensions of the Schr\"odinger equation,
such as the Gisin equation, provide a wave-function-based framework
for relaxation and its relation to classical Landau--Lifshitz-type
dynamics
\cite{gisinJPA81,gisinHelvPhysActa81,wieserEPJB15,wieserJPComm19}.
Recent studies have developed purity-preserving quantum equations of
magnetic damping \cite{Liu2024} and examined the influence of explicit
environments on entanglement and the emergence of classical magnetic
dynamics \cite{GarciaGaitanNikolic2024,uhrigNJP25}.

In previous work, related quantum--classical spin dynamics was studied
for spin dimers with Heisenberg and DM interactions and for a pure
DM spin trimer
\cite{WieserSanchezGalan2025Dimer,Wieser_Sanchez}.
Here we extend this framework to finite rings and ask how restoring
intersite quantum fluctuations changes local spin relaxation while the
microscopic spin remains $S=\tfrac12$.
This complements the semiclassical approach based on increasing $S$,
studied, for example, for Heisenberg trimers
\cite{mentrupPhysicaA00}.

We consider $N$ spin-$\tfrac12$ particles with nearest-neighbor
Heisenberg exchange, uniform DM interaction, and a magnetic field
parallel to the DM vector.
An interpolation parameter $\beta\in[0,1]$ connects the fully quantum
interaction at $\beta=1$ to a state-dependent mean-field generator at
$\beta=0$.
The intermediate regime provides a phenomenological way to vary the
weight of intersite fluctuation interactions at fixed microscopic spin;
$\beta$ is not identified with a microscopic decoherence rate.
The accompanying change in the overall interaction scale is accounted
for explicitly below.
Using dissipative Gisin--Schr\"odinger dynamics, we compare local spin
trajectories and relaxed states through local-moment magnitudes,
connected correlations, and single-site entropy.

For product initial states, the mean-field dynamics preserves the
product structure and the length of each local moment.
Restoring the quantum interaction permits entanglement and
time-dependent changes in local-moment magnitude.
The time traces and stationary observables reveal distinct even--odd
behavior. In the fully quantum limit, the even rings studied relax
toward nondegenerate ground states with vanishing local moments,
whereas the odd rings retain preparation-dependent moments within
their degenerate ground spaces.
Comparisons between independent initial states distinguish this
endpoint dependence from the robustness of the reported observables
at intermediate $\beta$.

As an analytical benchmark, we determine the finite-size saturation
threshold using the standard one-magnon analysis and an energy bound.
For sufficiently weak DM coupling, the quantum threshold remains
unchanged in even rings but varies linearly with $|D|$ in odd rings.
This identifies a specific consequence of chiral momentum selection
beyond the parity effect already present at $D=0$.
We also determine how the normalization of the interpolation rescales
this threshold.

The paper is organized as follows. Section~\ref{sect_theo} introduces the
Hamiltonian, the mean-field interpolation, and the dissipative dynamical
equation. Section~\ref{sec:analytic-solution} discusses the analytic
solution in the purely quantum regime, while
Section~\ref{sec:numerical} describes the numerical scheme used for
$\beta<1$. We then study the dependence of the stationary local moments,
connected correlations, and single-site entropy on $\beta$. Finally, we
derive the finite-size stability threshold of the fully polarized state
and compare the fully quantum and interpolating regimes.


\section{Theoretical Model}
\label{sect_theo}
We consider a system of $N$ spin-$\tfrac12$ sites located on the vertices of a regular $N$-gon and interacting only with their nearest neighbors. 

The Hilbert space of this system is $(\mathbb{C}^2)^{\otimes N}$. Throughout this paper, we set $\hbar=1$. As usual, we denote by $|\uparrow \; \rangle$ and $|\downarrow\; \rangle$ the eigenbasis of the single-spin operator $\sigma_z$. This induces the standard computational basis for the full Hilbert space. In this basis, the Hamiltonian of the system is a $2^N \times 2^N$ matrix and
each vector operator 
\begin{equation}
    \vec S_n = (S^x_n,S^y_n,S^z_n) \qquad n=1,\dots,N
\end{equation}
consists of three $2^N \times 2^N$ matrices which give the spin operator in each direction. For example, for $N=3$, the first site’s $y$-operator is $S_1^y=\tfrac12\bigl(\sigma_y \otimes I_{2} \otimes I_{2}\bigr)$.

We consider a Hamiltonian depending on a parameter $\beta\in[0,1]$,
which interpolates between a state-dependent mean-field description and
a fully quantum one:
\begin{equation} \label{eq:Hamiltonian}
  \hat H_\beta \;=\; \beta \hat H_{\rm QM} \;+\; (1-\beta) \hat H_{\rm MF} \;+\; \hat H_B \,.
\end{equation}
The first term is purely quantum mechanical and consists of an isotropic Heisenberg exchange and a Dzyaloshinsky--Moriya interaction.
\begin{equation} \label{eq:Ham_QM}
    \hat H_{\rm QM} \;=\; \hat H_J + \hat H_{\rm DM}.
\end{equation}
If we denote by $\langle nm\rangle$ the set of nearest neighbors,
\begin{equation}
    \langle nm\rangle := \{ (1,2), (2,3),\dots, (N-1,N),(N,1)\},
\end{equation}
the Heisenberg exchange contribution is
\begin{equation}
  \hat H_J \;=\; J \sum_{(n,m) \in \langle n m \rangle} \vec S_n \!\cdot\! \vec S_m.
\end{equation}
We focus throughout on the antiferromagnetic case $J>0$, where
frustration, even--odd effects, and connected spin correlations play a
central role. The exchange term is symmetric under interchange of $n$ and $m$ since spin operators acting on different sites commute. 

The antisymmetric Dzyaloshinsky--Moriya exchange interaction with a uniform coupling vector $\vec D$ is
\begin{equation}
   \hat H_{\rm DM} \;=\; \vec{D} \cdot  \sum_{(n,m) \in \langle n m \rangle}   \vec{S}_n \times \vec{S}_m .
\end{equation}
We take the DM vector $\vec D$ along the $z$-axis:
$\vec D = D (0,0,1)$. 

The mean-field contribution depends on the state of the system $|\psi\rangle$ and is given by
\begin{align} \label{eq:H_MF}
 \hat H_{\rm MF}(|\psi\rangle) \;=\; 
  \frac{1}{2} J \sum_{(n,m) \in \langle n m \rangle} 
  \Big[ \vec S_n \cdot \langle \vec S_m \rangle 
      + \vec S_m \cdot \langle \vec S_n \rangle \Big] \notag \\
      + \frac{1}{2} \vec{D}  \cdot \sum_{(n,m) \in \langle n m \rangle}  \Big[ \vec S_n \times \langle \vec S_m \rangle 
      - \vec S_m \times \langle \vec S_n \rangle \Big].
\end{align}
The spin expectation values are given by the standard quantum mechanical definition, 
$  \langle \vec S_n \rangle  \;=\;
   \langle\psi|\vec S_n|\psi\rangle$.

The first term is symmetric and the second antisymmetric, as are
their quantum counterparts.

The factors of $1/2$ in \eqref{eq:H_MF} are part of our
normalization convention. They ensure that the expectation value of
the mean-field operator reproduces the factorized interaction energy,
namely
\begin{equation}
\langle \hat H_{\rm MF}\rangle
=
\sum_{(n,m)\in\langle nm\rangle}
\left[
J\langle\vec S_n\rangle\cdot\langle\vec S_m\rangle
+
\vec D\cdot
\left(
\langle\vec S_n\rangle\times\langle\vec S_m\rangle
\right)
\right].
\end{equation}

The last contribution in the Hamiltonian is the Zeeman term, which couples the spins to an external magnetic field along the $z$-direction,
\begin{equation}
  \hat H_B \;=\; - B \sum_{n=1}^N S_n^z ,
\end{equation}
where $B$ denotes the Zeeman energy scale, $B=\gamma B_{\rm phys}$ with
$B_{\rm phys}$ the norm of the constant external magnetic field and $\gamma$ the gyromagnetic ratio. In this way, $B$, $J$, and $\vec D$ have dimensions of energy.

The dynamics of the system will be described by the Gisin--Schr\"odinger equation with damping constant $\alpha \ge 0$ \cite{wieserEPJB15,wieserJPComm19}:   
\begin{equation}
  \frac{d}{dt}\,|\psi(t)\rangle
  \;=\;
  -\,i \hat H\,|\psi(t)\rangle
  \;-\;\alpha\bigl(\hat H - \langle \hat H\rangle \bigr)\,|\psi(t)\rangle,
  \label{eq:gisin}
\end{equation}
where $\langle \hat H\rangle(t) = \langle \psi(t)|\hat H|\psi(t)\rangle$.
The dissipative term breaks unitarity while preserving the norm. For a
state-independent Hamiltonian it decreases the energy expectation; for the state-dependent model the decreasing quantity is instead the energy
functional
\begin{equation}
\mathcal E_\beta(\psi)
=
\beta\langle\hat H_{\rm QM}\rangle
+
\frac{1-\beta}{2}
\langle\hat H_{\rm MF}(\psi)\rangle
+
\langle\hat H_B\rangle ,
\label{eq:energy-functional-beta}
\end{equation}
as shown in Section~\ref{sec:numerical}.

The parameter $\beta$ interpolates between the state-dependent
mean-field generator and the fully quantum interaction. Specifically,
$\beta=0$ yields a purely mean-field description in which the spin
interactions are mediated through the instantaneous local expectation
values, whereas $\beta=1$ recovers the fully quantum Hamiltonian.

The meaning and normalization of the interpolation can be made explicit
without introducing a bath model. Let $I$ denote the identity on
$(\mathbb{C}^2)^{\otimes N}$. 
Write $\hat H_{\rm prod}=2\hat H_{\rm MF}-\langle \hat H_{\rm MF}\rangle \,I$ and
$\hat H_{\rm fl}=\hat H_{\rm QM}-\hat H_{\rm prod}$. Expanding each bond of
$\hat H_{\rm QM}$ in the fluctuations
$\delta\vec S_n=\vec S_n-\langle\vec S_n\rangle I$ produces a scalar, terms
linear in $\delta\vec S_n$, and terms bilinear in them; the first two are
exactly what $\hat H_{\rm prod}$ collects, so $\hat H_{\rm fl}$ is the sum of
the bond interactions bilinear in the fluctuations. Then
\begin{equation}
\begin{split}
\hat H_\beta&=\zeta\bigl(\hat H_{\rm prod}+\eta\,\hat H_{\rm fl}\bigr)+\hat H_B
 +\frac{1-\beta}{2} \langle \hat H_{\rm MF}\rangle \,I,\\[2pt]
\zeta&=\frac{1+\beta}{2},\qquad \eta=\frac{2\beta}{1+\beta}.
\end{split}
\label{eq:interpolation-meaning}
\end{equation}

The scalar term changes only the overall phase. With the normalization
adopted in \eqref{eq:H_MF}, the parameter $\beta$ therefore plays two
distinct roles, which \eqref{eq:interpolation-meaning} separates: $\eta$
weights the fluctuation interaction omitted by product-state factorization,
while $\zeta$ sets the overall interaction scale. At $B=0$ the latter can be
absorbed in $\tau=\zeta Jt$; at nonzero field $B/(\zeta J)$ must also be
specified. The same factor $\zeta$ rescales the linearized one-magnon
bandwidth, and with it the stability threshold of the fully polarized state,
as derived in Section~\ref{sec:critical-field}. 

The classical analogue of the Gisin--Schr\"odinger equation is the
Landau--Lifshitz--Gilbert equation. Indeed, for a spin-\(\tfrac12\) in a pure state with Hamiltonian
$\hat H=- \vec B\cdot \vec S$, the Gisin--Schr\"odinger evolution
induces Landau--Lifshitz-type dynamics, equivalent to Gilbert dynamics
after the usual parameter rescaling, for the magnetization vector
$\langle\psi(t)|\vec\sigma|\psi(t)\rangle$,
\cite{Wieser_Sanchez}.


\section{Dynamics of the system} 

\subsection{Quantum regime}
\label{sec:analytic-solution}

In the purely quantum regime, $\beta=1$, the Hamiltonian is
state-independent and the Gisin--Schr\"odinger equation can be solved
explicitly in the energy eigenbasis. Let
\begin{equation}
\hat H
=
\sum_k E_k\,|E_k\rangle\!\langle E_k|,
\qquad
\langle E_k|E_\ell\rangle=\delta_{k\ell},
\end{equation}
be a spectral decomposition of the Hamiltonian. Expanding the state as
\begin{equation}
|\psi(t)\rangle
=
\sum_k a_k(t)\,|E_k\rangle,
\qquad
a_k(0)=\langle E_k|\psi(0)\rangle,
\end{equation}
Eq.~\eqref{eq:gisin} gives a closed system of ordinary differential
equations for the amplitudes. It has the explicit solution
\begin{equation}
a_k(t)
=
\frac{
a_k(0)\,e^{(-iE_k-\alpha E_k)t}
}{
\sqrt{Z(t)}
},
\qquad
Z(t)
=
\sum_j |a_j(0)|^2e^{-2\alpha E_jt}.
\label{eq:ak-solution}
\end{equation}
The common factor $Z(t)^{-1/2}$ ensures that $\|\psi(t)\|=1$ for all times.

For a state-independent Hamiltonian, the energy expectation satisfies
\begin{align}
\frac{d}{dt}\langle\hat H\rangle
&=
-2\alpha
\left(
\langle\hat H^2\rangle
-
\langle\hat H\rangle^2
\right)
\nonumber\\
&=
-2\alpha\,\operatorname{Var}(\hat H)
\leq0.
\label{eq:energy-monotone}
\end{align}
Thus, for $\alpha>0$, the energy is non-increasing and decreases
strictly unless the state is supported entirely in a single energy
eigenspace.

Assume henceforth that $\alpha>0$ and define
\begin{equation}
E_\ast
=
\min\{E_k:\,a_k(0)\neq0\}.
\end{equation}
Then, as $t\to\infty$,
\begin{equation}
|a_k(t)|^2
\longrightarrow
\begin{cases}
\displaystyle
\frac{|a_k(0)|^2}
{\sum_{j:\,E_j=E_\ast}|a_j(0)|^2},
& E_k=E_\ast,\\[10pt]
0,
& E_k\neq E_\ast.
\end{cases}
\label{eq:long_time}
\end{equation}
Hence the state relaxes into the eigenspace corresponding to the
lowest energy having nonzero overlap with the initial state.

More precisely, let
\begin{equation}
P_\ast
=
\sum_{k:\,E_k=E_\ast}
|E_k\rangle\!\langle E_k|
\end{equation}
be the orthogonal projection onto the $E_\ast$-eigenspace. Then
\begin{equation}
\left\|
e^{iE_\ast t}|\psi(t)\rangle
-
\frac{
P_\ast|\psi(0)\rangle
}{
\|P_\ast|\psi(0)\rangle\|
}
\right\|
\longrightarrow0
\qquad
(t\to\infty).
\end{equation}

If the initial state has a nonzero projection onto the ground-state
subspace, then $E_\ast=E_{\min}$. If the ground state is
non-degenerate, the state approaches the ground state up to the global
phase $e^{-iE_{\min}t}$. If the ground-state energy is degenerate, the
state approaches, up to the same common phase, the normalized projection
of the initial state onto the ground-state subspace.

\subsection{Dynamics for $\beta <1$}
\label{sec:numerical}

For $\beta<1$, the instantaneous local moments make the equation nonlinear.
We use the normalized fourth-order Runge--Kutta (RK4) method, evaluating
the local moments and the state-dependent Hamiltonian at every RK4 stage.
The time step is $J\delta t=0.03$ and the damping is $\alpha=0.15$.
Stationary calculations are continued until
\begin{equation}
r=\frac{\|(\hat H_\beta-\langle \hat H_\beta\rangle)\psi\|}{J}<10^{-9}.
\label{eq:residual}
\end{equation}
A run that fails this criterion is not accepted. A fixed stopping time can terminate relaxation prematurely near slowly relaxing modes.
Selected runs at $(N,\beta)=(4,0.60),(5,0.50),(6,0.70),(6,0.75),(8,0.80)$
were repeated with half the time step; the changes in $M_\beta$, $C_{\rm NN}$
and $\overline S_{\rm VN}$ were below $6\times10^{-9}$.

The choice of relaxation equation is methodological. It retains precession,
acts on the full Hilbert space and preserves a pure state, so single-site
entropy measures entanglement. For $\beta=0$ and product initial states it
reduces to fixed-length Landau--Lifshitz-type spin dynamics. Imaginary-time
evolution is an alternative for finding stationary states but removes
precession. A Lindblad master equation is appropriate when a specified
environment and mixed states are required; Landau--Lifshitz damping can also
be obtained in that setting under local mean-field and weak-field assumptions
\cite{uhrigNJP25}. We use the Gisin equation as an effective deterministic
relaxation model, not as a microscopic description of environmental decoherence.
The dimensionless $\alpha=0.15$ keeps both precession and damping visible but is not fitted to a specific material. Repeating the first product
start for $N=5,6$ and $\beta=0,0.5,0.9$ with $\alpha=0.05$ and $0.5$ changed
the converged three observables by less than $4\times10^{-10}$.

On normalized states,
$\delta\langle\hat H_{\rm MF}\rangle
=4\operatorname{Re}\langle\delta\psi|\hat H_{\rm MF}|\psi\rangle$.
The factor $1/2$ in \eqref{eq:energy-functional-beta} therefore gives
$\delta\mathcal E_\beta
=2\operatorname{Re}\langle\delta\psi|\hat H_\beta|\psi\rangle$ and
\begin{equation}
\frac{d\mathcal E_\beta}{dt}
=-2\alpha\|(\hat H_\beta-\langle\hat H_\beta\rangle)\psi\|^2\leq0.
\label{eq:energy-general}
\end{equation}
Stationary rays are critical points of $\mathcal E_\beta$.

Let $\psi_*$ be a local minimum of $\mathcal E_\beta$ that is nondegenerate
modulo the phase and the continuous symmetries. Near $\psi_*$ the functional
satisfies
$\operatorname{Var}_{\psi}(\hat H_\beta)
\ge c\,[\mathcal E_\beta(\psi)-\mathcal E_\beta(\psi_*)]$
for some $c>0$, where
$\operatorname{Var}_{\psi}(\hat H_\beta)
=\|(\hat H_\beta-\langle\hat H_\beta\rangle)\psi\|^2$
is the quantity appearing in \eqref{eq:energy-general}. For $\alpha>0$, equation~\eqref{eq:energy-general} then gives local
exponential attraction to the stationary state or its symmetry orbit,
with common limiting values for observables constant on that orbit.

This local result does not exclude multiple attractors or preparation
dependence. At $\beta=1$, degeneracy of the lowest occupied eigenspace
can also preserve such dependence. We therefore compare independent
product initial states: for $\beta<1$, agreement of their stationary
observables supports robustness over the sampled preparations,
without proving uniqueness or convergence to the global minimum.

\subsection{Time-dependent comparison}
\label{sec:time-comparison}

\begin{figure}[!htbp]
\centering
\includegraphics[width=\linewidth]{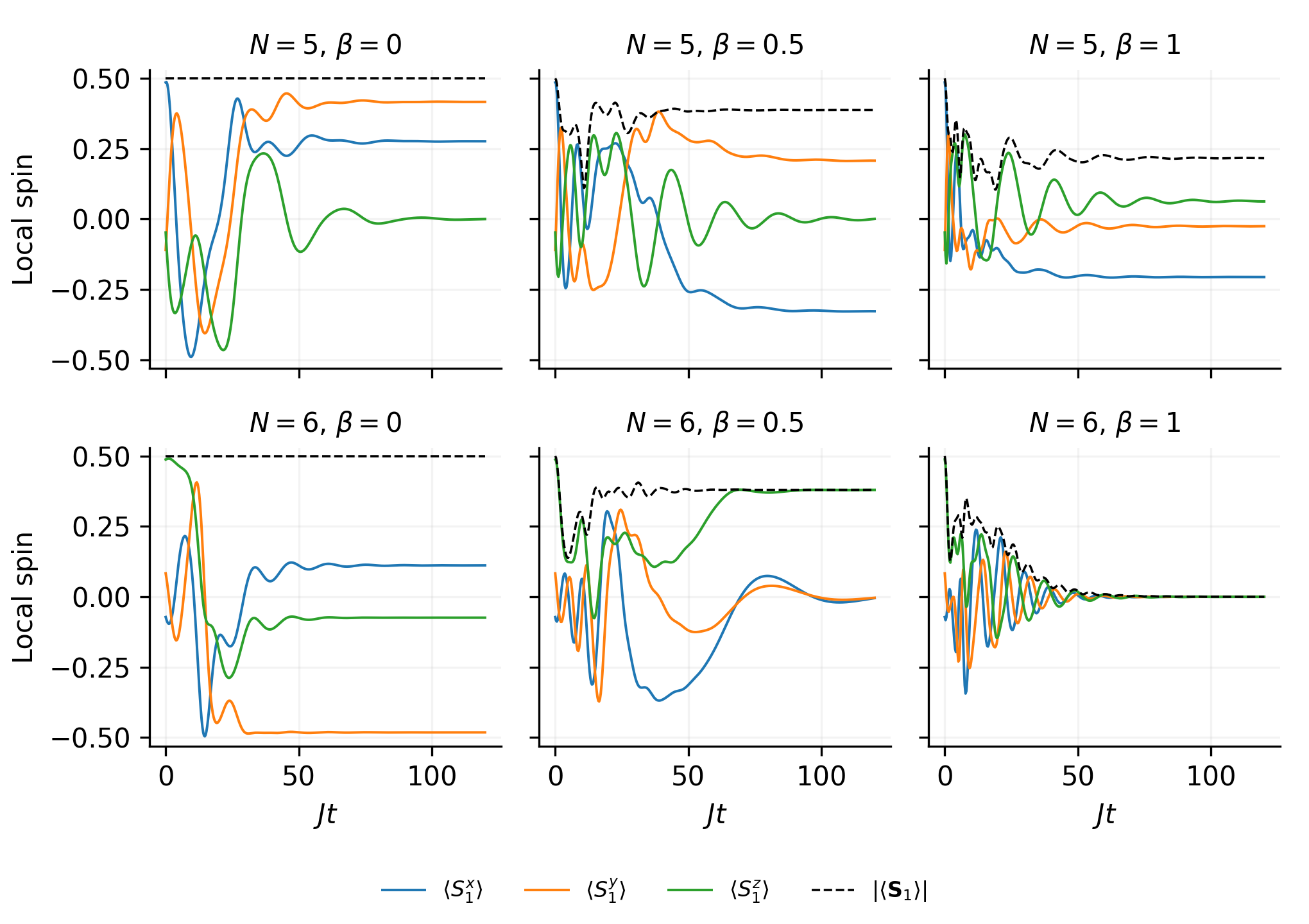}
\caption{Local spin components at site 1 for $N=5$ (top) and $N=6$ (bottom),
with $\beta=0,0.5,1$ from left to right. The dashed curve is the local-moment
length. Here $J=1$, $D=0.2$, $B=0$, $\alpha=0.15$ and $J\delta t=0.03$.
For each $N$, all three traces use the first product state generated with
seed $12345+1000N$. These are physical-time trajectories, not minimization
iterations.}
\label{fig:time-comparison}
\end{figure}

Figure~\ref{fig:time-comparison} compares $\beta=0,0.5,1$ for an odd and an
even ring, starting from the same random product state within each row.
All parameters and the physical time coordinate are common; the change in
interaction scale described by \eqref{eq:interpolation-meaning} is included in this comparison. At $\beta=0$, the local vector changes direction
at fixed length $1/2$. For $\beta>0$, entanglement can develop and its length
can decrease. In the fully quantum regime the even ring approaches a state
with vanishing local moments, whereas the odd ring retains a
preparation-dependent moment in its degenerate ground space.

\section{Quantum--classical correlation crossover}
\label{sec:crossover}

In the previous section, we studied the dynamical behavior of the system for a given interpolation parameter $\beta$.
We now analyze how this parameter controls the amount of quantum correlations present in the stationary states reached by the dynamics.

From the expression of the Hamiltonians \eqref{eq:Ham_QM} and \eqref{eq:H_MF}, 
\begin{equation}\label{eq:exp_H_MF}
   \langle \hat H_{\rm MF}\rangle =  \sum_{(n,m) \in \langle n m \rangle}  \big[ J
    \langle \vec S_n \rangle \cdot \langle \vec S_m \rangle 
   + \vec{D}  \cdot  (\langle \vec S_n \rangle \times \langle \vec S_m \rangle )\big],
\end{equation}
while the expectation value of the full quantum operator is
\begin{equation}
    \langle \hat H_{\rm QM} \rangle 
    = \sum_{(n,m) \in \langle n m \rangle}\big[ J
    \langle \vec S_n  \cdot  \vec S_m \rangle 
    +
     \vec{D} \cdot \langle \vec S_n  \times  \vec S_m  
   \rangle \big].
\end{equation}

The difference between these two expectation values is determined by the
connected (two-site) correlation functions of the spins,
\begin{align} \label{corre_funct}
C_{nm}
&=
\langle \vec S_n \cdot \vec S_m \rangle
-
\langle \vec S_n \rangle \cdot \langle \vec S_m \rangle \\
\label{corre_funct_DM}
\vec C_{nm}^{DM}
&=
\langle \vec S_n \times \vec S_m \rangle
-
\langle \vec S_n \rangle \times \langle \vec S_m \rangle.
\end{align}

Using these definitions one can write
\begin{equation}
\langle \hat H_{\rm QM}\rangle - \langle \hat H_{\rm MF}\rangle
=
\sum_{(n,m)\in\langle nm\rangle}
\Big[
J\,C_{nm} + \vec D\cdot \vec C_{nm}^{DM}
\Big],
\end{equation}
which shows that the connected correlations measure the correlation energy beyond the mean-field approximation.
For product states, these correlation functions vanish identically, thus $\langle \hat H_{\rm QM}\rangle = \langle \hat H_{\rm MF}\rangle$. The parameter $\beta$ controls the energetic weight of this
correlation energy and therefore governs the development of quantum correlations in the system.

For each site $n$, the operator $\vec S_n$ appears in the Hamiltonian
\eqref{eq:H_MF} in the two bonds $(n-1,n)$ and $(n,n+1)$.
Collecting these contributions, the mean-field Hamiltonian can be written as
\begin{equation}
\hat H_{\rm MF}(|\psi\rangle)
=
\sum_{n=1}^N \vec S_n\cdot \vec B_n^{\rm eff},
\end{equation}
where
\begin{equation}
\vec B_n^{\rm eff}
=
\frac{J}{2}
\bigl(
\langle \vec S_{n-1}\rangle+\langle \vec S_{n+1}\rangle
\bigr)
+
\frac12
\bigl(
\langle \vec S_{n+1}\rangle-\langle \vec S_{n-1}\rangle
\bigr)\times \vec D ,
\end{equation}
with indices understood modulo $N$.

In the pure mean-field limit $\beta=0$, the Hamiltonian contains only single-spin operators coupled to the effective classical fields $\vec B_n^{\rm eff}$. For product initial states, the dynamics remains confined to the product-state space and is driven toward self-consistent stationary states. In this limit,
\eqref{eq:energy-functional-beta} reduces to
\[
\mathcal E_0(\psi)
=
\frac12\langle\hat H_{\rm MF}(\psi)\rangle
+
\langle\hat H_B\rangle.
\]
Since this functional depends only on local expectation values,
connected correlations cannot provide an additional energetic lowering.
By contrast, the genuine two-spin interactions present as
$\beta\to1$ allow connected correlations to lower the energy beyond
the factorized approximation.

We use $D/J=0.2$ as a moderate illustrative coupling, without fitting a specific material.

\begin{figure}[!htbp]
\centering
\includegraphics[width=0.7\linewidth]{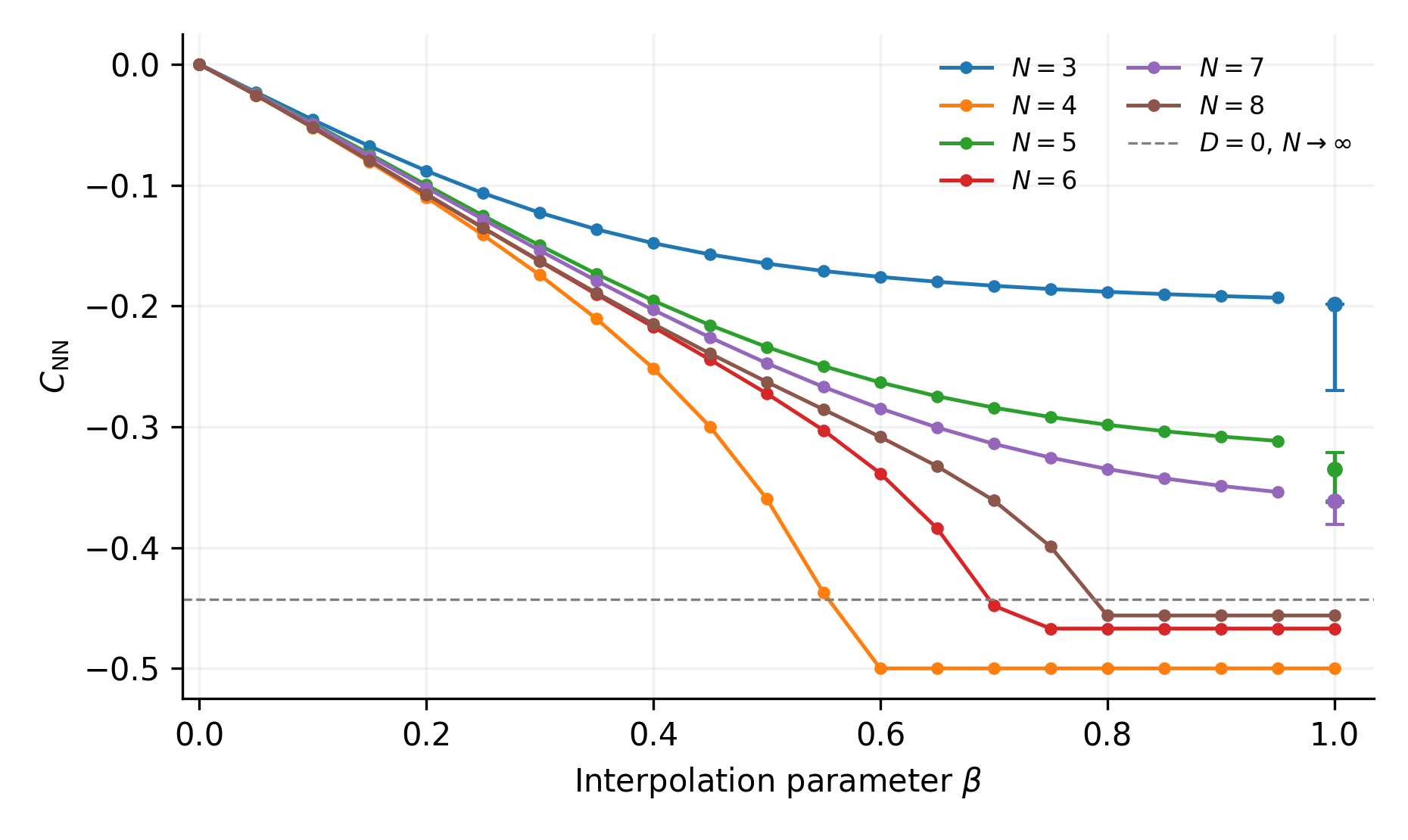}
\caption[Average nearest-neighbor connected correlation as a function of $\beta$]{
Average nearest-neighbor connected correlation $C_{\mathrm{NN}}$ as a function
of the interpolation parameter $\beta$ for rings with
$N=3,4,5,6,7,8$ spins. For small values of $\beta$, the connected nearest-neighbor correlations
are close to zero, consistent with the mean-field description.
As $\beta$ increases, the magnitude of the correlations grows and
$C_{\rm NN}$ becomes increasingly negative, reflecting the development of
antiferromagnetic quantum correlations. The even rings generally develop
stronger negative correlations, while the odd rings are affected by
geometrical frustration. At $\beta=1$, odd-ring points use the first
initial state and vertical bars show the range across all four starts;
these are preparation ranges, not statistical error bars. The dashed
reference is for $D=0$ and $N\to\infty$.}
\label{fig:aver_corr}
\end{figure}

If we neglect the Dzyaloshinsky--Moriya interaction and the external
magnetic field, the spin-$\tfrac12$ antiferromagnetic Heisenberg chain has the exact Bethe ansatz nearest-neighbor correlation
\cite{Bethe1931,Hulthen1938,Auerbach1994}
\begin{equation}
\langle \vec S_n \cdot \vec S_{n+1} \rangle
=
\frac{1}{4}-\ln 2
\approx -0.443
\end{equation}
in the thermodynamic limit. Since the ground state is
$SU(2)$-symmetric and translationally invariant, one has
$\langle \vec S_n\rangle=0$. Therefore, the average nearest-neighbor
connected correlation
\begin{equation}
C_{\rm NN}
:=
\frac{1}{N}
\sum_{(n,m)\in\langle nm\rangle}
\left(
\langle \vec S_n\cdot \vec S_m\rangle
-
\langle \vec S_n\rangle\cdot\langle \vec S_m\rangle
\right)
\end{equation}
satisfies
\begin{equation}
\lim_{N\to\infty} C_{\rm NN}
=
\frac14-\ln 2 .
\end{equation}
The finite rings considered here need not attain this value because of
finite-size effects, odd-$N$ frustration, and the nonzero
Dzyaloshinsky--Moriya interaction used in the simulations; see Fig.~\ref{fig:aver_corr}.

This crossover is also reflected in the behavior of the local
magnetization. In the following, ``average local magnetization'' refers
to the site average of the magnitudes of the local spin expectation
values, and should not be confused with the net magnetization of the
ring. We define
\begin{equation}
M_\beta(t)
=
\frac{1}{N}\sum_{n=1}^N |\langle \vec S_n \rangle(t)|.
\end{equation}
The expectation values $\langle \vec S_n\rangle(t)$ depend on the evolving
state $|\psi(t)\rangle$ and therefore, in principle, on the initial
condition. 

\begin{figure}[!htbp]
\centering
\includegraphics[width=0.7\linewidth]{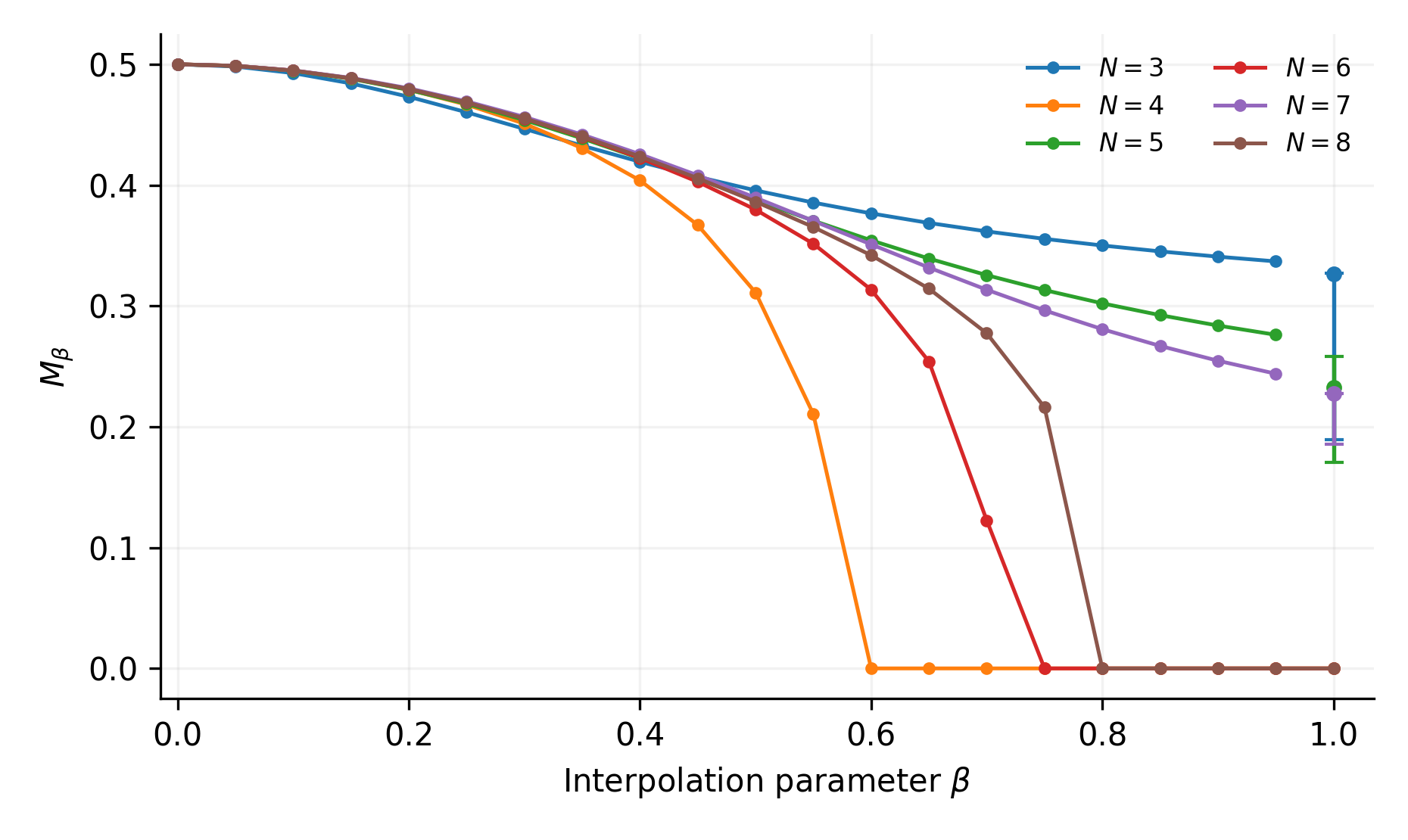}
\caption[Average local magnetization as a function of $\beta$]{
Average local magnetization
$M_\beta$ as a function of the interpolation parameter $\beta$
for rings with $N=3,4,5,6,7,8$ spins. The magnetization generally decreases as $\beta$ increases, reflecting the suppression of classical local spin order in the correlated quantum regime. The sharper drops observed for
even $N$ reflect the absence of geometrical frustration, while the odd
rings retain a larger residual magnetization because a perfect alternating
antiferromagnetic arrangement is incompatible with the periodic boundary
condition. Odd-ring endpoints and bars follow the convention in
Fig.~\ref{fig:aver_corr}; they need not equal the one-sided limit
$\beta\to1^-$.}
\label{fig:magn}
\end{figure}

The quantity plotted in Fig.~\ref{fig:magn} corresponds to the average local magnetization after relaxation under the Gisin dynamics:
\begin{equation}
M_\beta
:=
\lim_{t\to\infty}
\frac{1}{N}\sum_{n=1}^N |\langle \vec S_n\rangle(t)|.
\end{equation}

By the local attraction established in Section~\ref{sec:numerical}, this limit
exists whenever the relaxation reaches such a stationary state, since
$|\langle\vec S_n\rangle|$ is constant on the symmetry orbit.

The numerical parameters are $J=1$, $D=0.2$, $B=0$ and $\alpha=0.15$.
For each of the 21 values $\beta=0,0.05,\ldots,1$ and each $N=3,\ldots,8$,
we retain the four random product starts and seed rule of the original
notebook. Each run is now continued to \eqref{eq:residual}; the required
times range from $Jt=150$ to $21030$. For $\beta<1$ we plot the converged
run with the lowest $\mathcal E_\beta$. Across the four starts, the maximum
spread in $M_\beta$ is below $5\times10^{-8}$ on this grid, and the spreads
in $C_{\rm NN}$ and $\overline S_{\rm VN}$ are below $5\times10^{-9}$.
These comparisons support robustness of the reported observables within
the sampled preparations; they do not certify global minimization or
exclude other attractors.

The fully quantum endpoint requires a separate convention. At $B=0$,
the unique ground states found for $N=4,6,8$ have vanishing local moments
by time-reversal symmetry. For $N=3,5,7$, ground-state degeneracy permits
different local moments at the same energy. As the exact solution in
Section~\ref{sec:analytic-solution} shows, relaxation preserves the normalized
projection of the initial state onto that ground space. The four starts give
$M_1\in[0.1890,0.3270]$, $[0.1706,0.2583]$ and $[0.1857,0.2278]$ for
$N=3,5,7$, respectively, while their energies agree within $2\times10^{-15}J$.
At $\beta=1$ we therefore plot the first start deterministically and show the
full preparation range. Energy alone cannot select a unique endpoint.

\subsection{Single-site entropy}

To further characterize the quantum--classical crossover, we compute the
single-site von Neumann entropy. Since the Gisin--Schr\"odinger evolution
is formulated at the wave-function level, the global state remains pure,
$|\psi\rangle \in (\mathbb C^2)^{\otimes N}$. Tracing out all spins
except site $n$ gives the reduced density matrix
\begin{equation}
\rho_n
=
\operatorname{Tr}_{\{1,\dots,N\}\setminus\{n\}}
\left(
|\psi\rangle\langle\psi|
\right),
\end{equation}
and the corresponding single-site entropy is
\begin{equation}
S_{\rm VN}^{(n)}
=
-\operatorname{Tr}\left(\rho_n\log_2\rho_n\right).
\end{equation}
For a pure global state, $S_{\rm VN}^{(n)}$ measures the entanglement
between site $n$ and the remaining $N-1$ spins. It vanishes precisely
when site $n$ is unentangled from the rest of the ring.

For a spin-$\tfrac12$ system, the $2\times2$ reduced density matrix is
completely determined by the local spin expectation value. Let
$\vec\sigma=(\sigma_x,\sigma_y,\sigma_z)$ denote the vector of Pauli
matrices acting on the reduced single-site Hilbert space. Then
\begin{equation}
\rho_n
=
\frac12 I_2
+
\langle \vec S_n\rangle\cdot \vec\sigma .
\end{equation}
The eigenvalues of $\rho_n$ are therefore
\begin{equation}
\lambda_\pm
=
\frac12
\pm
|\langle \vec S_n\rangle|,
\end{equation}
and the entropy can be written as
\begin{equation}
S_{\rm VN}^{(n)}
=
-\lambda_+\log_2\lambda_+
-\lambda_-\log_2\lambda_- .
\end{equation}
Thus the single-site entropy is a monotone decreasing function of the
local magnetization amplitude $|\langle \vec S_n\rangle|$: it is zero
for a locally pure spin state with $|\langle \vec S_n\rangle|=1/2$, and maximal when the local spin expectation value vanishes.

\begin{figure}[!htbp]
\centering
\includegraphics[width=0.7\linewidth]{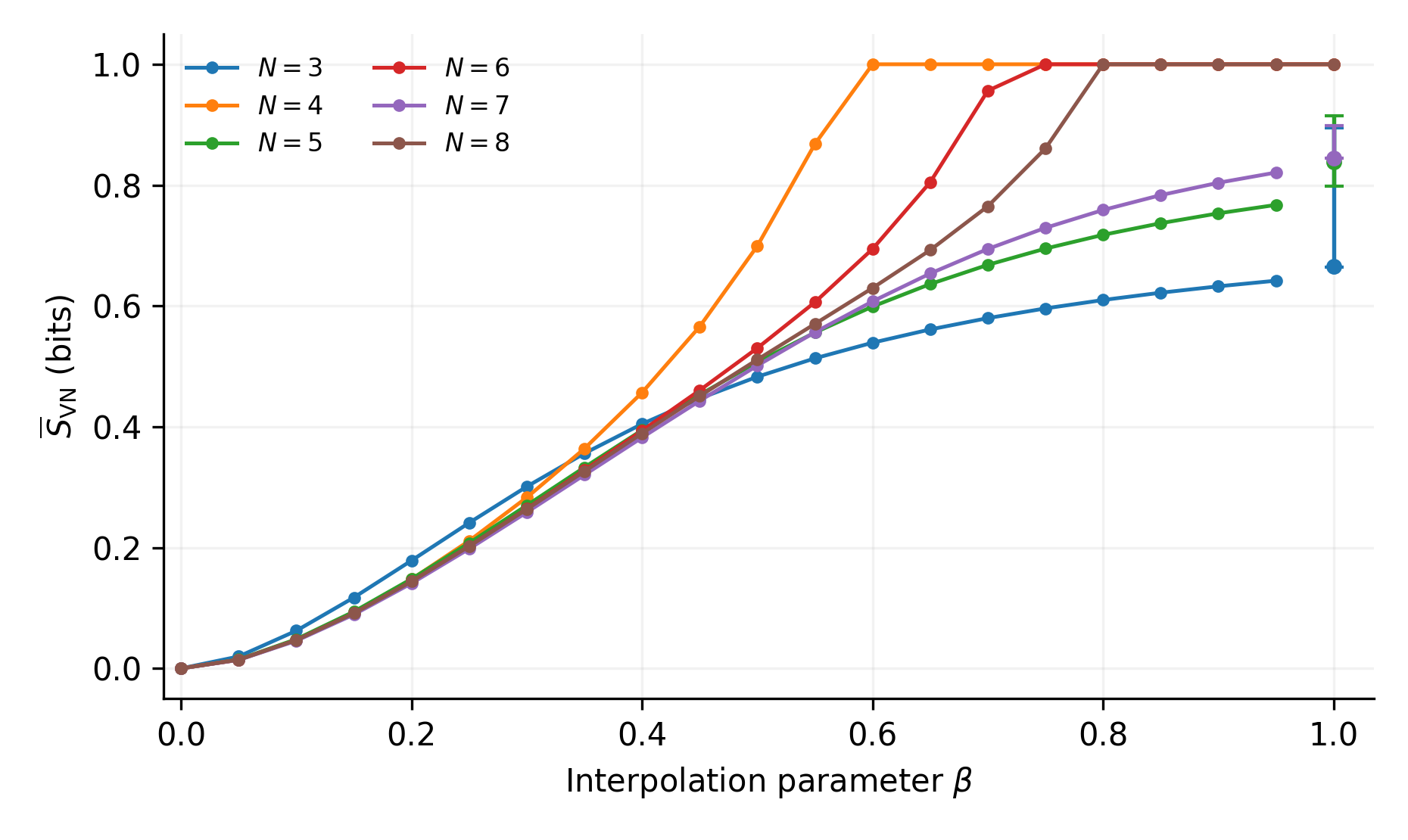}
\caption{Average single-site von Neumann entropy
$\overline S_{\rm VN} =
\frac1N
\sum_{n=1}^N S_{\rm VN}^{(n)}$ as a function of the interpolation parameter
$\beta$ for rings with $N=3,4,5,6,7,8$. The entropy is small in the
mean-field regime and increases as the dynamics approaches the correlated
quantum regime. The even rings show the largest entropy increase,
consistent with the stronger suppression of their local magnetization.
Odd-ring endpoints and preparation ranges use the convention of
Fig.~\ref{fig:aver_corr}.}
\label{fig:entropy-vs-beta}
\end{figure}

As shown in Fig.~\ref{fig:entropy-vs-beta}, the mean-field regime relaxes
toward states close to product states and therefore has small single-site
entropy. In the correlated quantum regime, genuine two-spin interactions
suppress the local magnetic moments and increase the entanglement between
individual spins and the rest of the ring. The even rings show the
strongest entropy increase, consistent with their sharper loss of local
magnetization, whereas the odd rings retain a larger residual local moment because of geometrical frustration.

\paragraph{Role of the DM interaction.}
Parity effects are already present at $D=0$; DM exchange does not create
geometrical frustration. A matched check at $\beta=0.5$ illustrates its
additional effect on the relaxed observables:
\begin{center}
\begin{tabular}{ccrr}
\hline
$N$ & $D/J$ & $M_{0.5}$ & $C_{\rm NN}$\\
\hline
5 & 0 & 0.376992 & $-0.244125$\\
5 & 0.2 & 0.387596 & $-0.234001$\\
6 & 0 & 0.377631 & $-0.278334$\\
6 & 0.2 & 0.379715 & $-0.272492$\\
\hline
\end{tabular}
\end{center}
The same four preparations and convergence criterion were used for each
comparison. Thus the even--odd contrast survives at $D=0$, while DM exchange
changes the local moments and correlations. Its chiral momentum selection
is especially transparent in the field threshold derived next.

\section{Critical magnetic field}
\label{sec:critical-field}

The purpose of this section is twofold. First, we recall the standard
one-magnon derivation of the saturation field for the finite
spin-$\tfrac12$ Heisenberg ring and extend the calculation to include
a uniform Dzyaloshinsky--Moriya interaction. Second, we apply the same
stability argument to the $\beta$-dependent interpolating Hamiltonian.
The one-magnon technique and the $D=0$ parity effect are standard. The purpose
here is to identify explicitly the finite-ring DM dependence and to separate
the interpolation's normalization factor from its effect on correlations.

With the normalization adopted in \eqref{eq:H_MF}, the full quantum
interaction produces the complete transverse one-magnon hopping,
whereas the linearization of the state-dependent mean-field term around
the fully polarized state produces one half of that hopping. Since the
two contributions enter the interpolating Hamiltonian with weights
$\beta$ and $1-\beta$, respectively, the effective one-magnon bandwidth
is multiplied by
\begin{equation}
\beta+\frac{1-\beta}{2}
=
\frac{1+\beta}{2}.
\end{equation}
Consequently, the finite-size stability threshold of the fully
polarized state is rescaled by the same factor.
\subsection{Purely quantum case}

In the purely quantum case $\beta=1$, the saturation field is obtained
from the one-magnon instability of the fully polarized state. This is the
standard saturation-field mechanism for the antiferromagnetic
spin-$\tfrac12$ Heisenberg chain. Related finite-size and frustrated-spin magnetization phenomena, including
magnetization plateaus, have been studied extensively in quantum spin
systems \cite{OshikawaYamanakaAffleck1997,HoneckerSchulenburgRichter2004}. We now recall the elementary
finite-ring calculation, which will later be adapted to the
$\beta$-dependent interpolating model.

We denote the fully polarized state by
\begin{equation}
|F\rangle = |\uparrow \uparrow \cdots \uparrow\rangle .
\end{equation}
The Hamiltonian commutes with the total magnetization
\begin{equation}
S^z_{\rm tot} = \sum_{n=1}^N S_n^z ,
\end{equation}
so the Hilbert space decomposes into invariant magnetization sectors
labelled by the eigenvalue $m$ of $S^z_{\rm tot}$.

Within each sector, the Zeeman term acts as a constant shift.
Consequently, the lowest energy in the sector with magnetization $m$
depends linearly on the magnetic field:
\begin{equation}
E(m,B) = E_0(m) - Bm,
\end{equation}
where $E_0(m)$ denotes the lowest eigenvalue of the field-free
Hamiltonian $\hat H_{\rm QM}$ restricted to the sector with magnetization
$m$. The transition to the fully polarized phase occurs when the branch
with maximal magnetization $m=N/2$ intersects the adjacent one-magnon
sector $m=N/2-1$. Hence the saturation field is
\begin{equation}
B_c(N)
=
E_0\left(\frac{N}{2}\right)
-
E_0\left(\frac{N}{2}-1\right).
\end{equation}

We first consider the well-known case without Dzyaloshinsky--Moriya interaction,
$D=0$.

In the fully polarized sector, each bond contributes $J/4$, and therefore
\begin{equation}
E_0\left(\frac{N}{2}\right)
=
\frac{JN}{4}.
\end{equation}

In the one-magnon sector, a convenient basis is
\begin{equation}
|n\rangle = S_n^- |F\rangle,
\end{equation}
where the spin at site $n$ is flipped down. By translation invariance, the eigenstates of the Hamiltonian are plane waves
\begin{equation}
|k\rangle
=
\frac{1}{\sqrt N}
\sum_{n=1}^{N}
e^{ikn}|n\rangle,
\qquad
k=\frac{2\pi q}{N},
\qquad
q=0,\dots,N-1.
\end{equation}

A direct computation gives the field-free one-magnon dispersion
\begin{equation}
E(k)
=
J\left(\frac{N}{4}-1+\cos k\right).
\end{equation}
If $N$ is even, the
momentum $k=\pi$ is allowed, while for $N$ odd, the momentum $k=\pi$ is not allowed and the lowest energy is achieved for 
$k_\pm=\pi\pm\frac{\pi}{N}$. Hence,
\begin{equation}
B_c^{(0)}(N)
=
\begin{cases}
2J, & N\ \text{even},\\[6pt]
J\left(1+\cos\left(\dfrac{\pi}{N}\right)\right), & N\ \text{odd}.
\end{cases}
\label{eq:Bc_cases}
\end{equation}

We now include a uniform Dzyaloshinsky--Moriya interaction parallel to the external magnetic field:
\begin{equation}
\hat H
=
J \sum_{n=1}^{N} \vec S_n\cdot \vec S_{n+1}
+
D \sum_{n=1}^{N}(\vec S_n\times \vec S_{n+1})_z
-
B \sum_{n=1}^{N} S_n^z, \qquad
 \vec S_{N+1}\equiv \vec S_1.
\end{equation}

A uniform Dzyaloshinsky--Moriya interaction in a one-dimensional
spin chain can be gauged out of the bulk transverse exchange by a
site-dependent rotation about the DM axis. The resulting bulk
Hamiltonian is of XXZ type, while for a finite periodic ring the
accumulated rotation generally reappears as a twisted boundary
condition \cite{PerkCapel1976}.
For this reason, in the present finite-ring calculation, it is more convenient to
retain the original spin variables and derive the one-magnon dispersion directly.

The fully polarized state is unaffected by the Dzyaloshinsky--Moriya term, since the latter vanishes on $|F\rangle$. In the one-magnon sector, however, the
DM interaction contributes an antisymmetric
hopping term. The field-free one-magnon dispersion is
\begin{equation}
E(k)
=
J\left(\frac{N}{4}-1+\cos k\right)
+
D\sin k.
\end{equation}
The saturation field is therefore
\begin{equation}
B_c(N)
=
\max_{k\in \frac{2\pi}{N}\mathbb Z}
\left[
J(1-\cos k)-D\sin k
\right].
\label{eq:Bc_DM_finite}
\end{equation}

This expression can also be written in closed even--odd form. Let
\begin{equation}
R=\sqrt{J^2+D^2},
\qquad
\phi=\arctan\left(\frac{D}{J}\right),
\end{equation}
and let $\operatorname{dist}(x,A)$ denote the shortest distance
on the circle $\mathbb R/2\pi\mathbb Z$ from $x$ to the set $A$. Since
\begin{equation}
J(1-\cos k)-D\sin k
=
J-R\cos(k-\phi),
\end{equation}
the maximum is obtained by choosing an allowed momentum $k$ as close as
possible to $\pi+\phi$. Therefore
\begin{equation}
B_c(N)
=
\begin{cases}
\displaystyle
J+R\cos\!\left[
\operatorname{dist}
\left(
\phi,\frac{2\pi}{N}\mathbb Z
\right)
\right],
& N\ \mathrm{even},\\[16pt]
\displaystyle
J+R\cos\!\left[
\operatorname{dist}
\left(
\phi,\frac{2\pi}{N}\left(\mathbb Z+\frac12\right)
\right)
\right],
& N\ \mathrm{odd}.
\end{cases}
\label{eq:Bc_DM_even_odd}
\end{equation}
For $D=0$, one has $\phi=0$ and $R=J$, so one recovers \eqref{eq:Bc_cases}.
More explicitly, while $|D|/J\leq\tan(\pi/N)$ an even ring retains
$B_c=2J$. For an odd ring, the nearest pair of momenta gives
\begin{equation}
B_c=J[1+\cos(\pi/N)]+|D|\sin(\pi/N),
\qquad |\phi|\leq2\pi/N.
\label{eq:weak-dm}
\end{equation}
Thus a weak DM coupling leaves the even-ring threshold unchanged but
changes the odd-ring threshold linearly. This distinguishes the role of DM
from the parity effect already present without it.

In the thermodynamic limit, the allowed momenta become dense, and the
minimum value of $\cos(k-\phi)$ is $-1$. Hence
\begin{equation}
\lim_{N\to\infty} B_c(N)
=
J+\sqrt{J^2+D^2}.
\end{equation}
For $D=0$, this reduces to $2J$, the usual saturation field of the antiferromagnetic Heisenberg chain. The result is not
$2\sqrt{J^2+D^2}$, because the Dzyaloshinsky--Moriya term renormalizes only the transverse magnon hopping, whereas the longitudinal exchange contribution $J S_n^zS_{n+1}^z$ remains governed by $J$. 

\subsection{Effect of $\beta$ on the saturation field}

For the interpolating model, the Hamiltonian becomes state-dependent
whenever $\beta<1$, through $\hat H_{\rm MF}(|\psi\rangle)$. Hence the
saturation field is most naturally defined as a stability threshold. We
define $B_c(\beta;N)$ as the value of the external field at which the
fully polarized fixed point
\begin{equation}
|F\rangle=|\uparrow\cdots\uparrow\rangle
\end{equation}
changes linear stability with respect to one-magnon perturbations.
Equivalently, $B_c(\beta;N)$ is the value of the field at which the
lowest linearized one-magnon gap closes. For $B>B_c(\beta;N)$, all
one-magnon gaps are positive and the fully polarized state is linearly
stable against these perturbations. For $B<B_c(\beta;N)$, at least one one-magnon mode lowers the
energy, and the fully polarized state becomes unstable.

We first consider $D=0$. Linearizing around the fully polarized state, we
write
\begin{equation}
|\psi\rangle
=
|F\rangle
+
\sum_{n=1}^{N} b_n S_n^-|F\rangle
+
O(\|b\|^2).
\end{equation}
To first order,
\begin{equation}
\langle S_n^z\rangle
=
\frac12+O(\|b\|^2),
\qquad
\langle S_n^+\rangle
=
b_n+O(\|b\|^2).
\end{equation}

At $|F\rangle$, the frozen mean-field Hamiltonian is purely longitudinal:
\begin{equation}
\hat H_{\rm MF}(|F\rangle)
=
\frac{J}{2}\sum_{n=1}^{N} S_n^z .
\end{equation}
However, when the state dependence of $\hat H_{\rm MF}$ is linearized
around $|F\rangle$, transverse terms appear and generate nearest-neighbor hopping of the one-magnon amplitudes. With the normalization used in
\eqref{eq:H_MF}, the mean-field contribution gives one half of the one-magnon hopping produced by the full quantum exchange term. Therefore the linearized one-magnon gap is
\begin{equation}
\Delta_\beta(k)
=
B
-
\frac{1+\beta}{2}J(1-\cos k).
\end{equation}

The saturation field is obtained when the minimum of $\Delta_\beta(k)$
vanishes. Hence
\begin{equation}
B_c^{(0)}(\beta;N)
=
\frac{1+\beta}{2}J
\begin{cases}
2, & N\ \text{even},\\[6pt]
1+\cos\left(\dfrac{\pi}{N}\right), & N\ \text{odd}.
\end{cases}
\label{eq:Bc_beta_D0}
\end{equation}
For $\beta=1$, this reproduces the purely quantum result
\eqref{eq:Bc_cases}.

We now include a uniform Dzyaloshinsky--Moriya interaction. Linearizing
around the fully polarized state gives a complex nearest-neighbor hopping amplitude. Equivalently, the one-magnon gap becomes
\begin{equation}
\Delta_\beta(k)
=
B
-
\frac{1+\beta}{2}
\left[
J(1-\cos k)-D\sin k
\right].
\end{equation}
Thus the finite-size saturation field is
\begin{equation}
B_c(\beta;N)
=
\frac{1+\beta}{2}
\max_{k\in \frac{2\pi}{N}\mathbb Z}
\left[
J(1-\cos k)-D\sin k
\right].
\label{eq:Bc_beta_DM_general}
\end{equation}
For $\beta=1$, this reduces to the purely quantum finite-size result
\eqref{eq:Bc_DM_finite}.
Using \eqref{eq:Bc_DM_even_odd}, this may be written as
\begin{equation}
B_c(\beta;N)
=
\frac{1+\beta}{2}
\begin{cases}
\displaystyle
J+R\cos\!\left[
\operatorname{dist}
\left(
\phi,\frac{2\pi}{N}\mathbb Z
\right)
\right],
& N\ \mathrm{even},\\[16pt]
\displaystyle
J+R\cos\!\left[
\operatorname{dist}
\left(
\phi,\frac{2\pi}{N}\left(\mathbb Z+\frac12\right)
\right)
\right],
& N\ \mathrm{odd},
\end{cases}
\label{eq:Bc_beta_DM_even_odd}
\end{equation}
where
\begin{equation}
R=\sqrt{J^2+D^2},
\qquad
\phi=\arctan\left(\frac{D}{J}\right).
\end{equation}

In the thermodynamic limit, this becomes
\begin{equation}
B_c(\beta)
=
\frac{1+\beta}{2}
\left(
J+\sqrt{J^2+D^2}
\right),
\qquad
N\to\infty.
\end{equation}


For completeness, the threshold is also a global saturation threshold for
the stated energy functional. In the quantum model the hard-core-boson form
relative to $|F\rangle$ is
$\hat H-E_F=\sum_{nm}h_{nm}b_n^\dagger b_m+J\sum_n n_n n_{n+1}$.
Here $h$ is the one-magnon matrix. At $B\geq B_c(N)$ it is positive
semidefinite, and the density interaction is nonnegative since $J>0$.
Thus no multi-magnon state lowers the energy before the one-magnon gap closes.

For the interpolation set $b=B/\zeta$ and let
$\rho_{\rm prod}=\bigotimes_n\rho_n$ be the product of the one-site reduced
states of $\psi$. Since its interaction expectation equals $\langle \hat H_{\rm MF}\rangle$,
\begin{equation}
\frac{\mathcal E_\beta(\psi)}{\zeta}
=\eta\langle \hat H_{\rm QM}-bS^z_{\rm tot}\rangle_\psi
 +(1-\eta)\operatorname{Tr}[\rho_{\rm prod}(\hat H_{\rm QM}-bS^z_{\rm tot})].
\label{eq:saturation-bound}
\end{equation}
For $b\geq B_c(N)$ both expectations are bounded below by the polarized
quantum ground energy, and $|F\rangle$ attains that bound. Below it a small
product-state canting in the unstable wavevector lowers the functional.
Hence \eqref{eq:Bc_beta_DM_general} is exact for this functional; at equality
the polarized state is a global minimizer, not necessarily a unique one.
The ratio $B_c(\beta;N)/\zeta$ is independent of $\beta$, so this rescaling
should not be interpreted as an independent measure of quantum correlations.

\section{Conclusion}

We have studied a finite spin-$\tfrac12$ ring with Heisenberg exchange,
Dzyaloshinsky--Moriya interaction, and an external magnetic field through a quantum--mean-field interpolation controlled by the parameter $\beta$.
Using dissipative Gisin--Schr\"odinger dynamics, we found that increasing $\beta$ enhances connected antiferromagnetic correlations, suppresses the average local-moment magnitude, and increases the single-site entropy. The numerical results display a clear even--odd finite-size effect: even rings can relax
toward more strongly correlated states with smaller local moments, whereas odd rings retain larger local moments through geometrical
frustration, becoming preparation-dependent within their
degenerate ground spaces in the fully quantum limit.

We also derived the finite-size saturation field from the one-magnon instability of the fully polarized state. A uniform Dzyaloshinsky--Moriya interaction shifts the optimal one-magnon momentum, so the finite-ring critical field depends on the discrete allowed momenta. In the interpolating model, the parameter $\beta$ rescales the linearized one-magnon bandwidth by the factor $(1+\beta)/2$, yielding an explicit
formula for $B_c(\beta;N)$. These results extend the previous dimer and trimer analyses to finite rings and show how quantum correlations, geometrical frustration, and finite-size momentum quantization enter the
quantum--classical crossover. 


\section*{Code availability}

The explanatory notebook and supporting files used to generate the data and
figures in this paper are publicly available in the
\href{https://github.com/raulsanchezgalan/quantum-classical-spin-rings}{GitHub repository associated with this work}.

\bibliographystyle{SciPost_bibstyle}
\bibliography{references}

\end{document}